\documentclass[useAMS,usenatbib]{mn2e}
\usepackage{graphicx}
\usepackage{color}
\usepackage{epsfig}
\usepackage{amsmath,amssymb}
\usepackage{natbib}

\def\simge{
    \mathrel{\rlap{\raise 0.511ex
        \hbox{$>$}}{\lower 0.511ex \hbox{$\sim$}}}}
\def\simle{
    \mathrel{\rlap{\raise 0.511ex
        \hbox{$<$}}{\lower 0.511ex \hbox{$\sim$}}}}

\newcommand{\figref}[1]{Figure~\ref{#1}}

\newcommand{\secref}[1]{Section~\ref{#1}}

\newcommand{\msun}{M_{\odot}}

\newcommand{\msunyr}{M_\odot~{\rm yr}^{-1}}
\newcommand{\mdot}{\dot{M}_*}

\newcommand{\hii}{H{\sc ii} }
\newcommand{\yr}{{\rm yr} }
\newcommand{\Myr}{{\rm Myr} }

\title[SMS formation by realistic episodic accretion]
{Supermassive star formation via episodic accretion: 
protostellar disc instability and radiative feedback efficiency}

\author[Y. Sakurai, et al.]
{Y. Sakurai$^{1}$, 
E. I. Vorobyov$^{2,3}$, T. Hosokawa$^{1,4}$,
N. Yoshida$^{1,5}$, K. Omukai$^{6}$
\newauthor
and H. W. Yorke$^{7}$
\\
$^{1}$Department of Physics, School of Science, University of Tokyo, 
Bunkyo, Tokyo 113-0033, Japan\\
$^{2}$Department of Astrophysics, The University of Vienna, Vienna, 1180, Austria\\
$^{3}$Research Institute of Physics, Southern Federal University, Rostov-on-Don 344090, Russia\\
$^{4}$Research Center for the Early Universe, 
University of Tokyo, Bunkyo, Tokyo 113-0033, Japan\\
$^{5}$Kavli Institute for the Physics and Mathematics 
of the Universe (WPI), Todai Institutes for Advanced Study, \\
the University of Tokyo, Kashiwa, Chiba 277-8583, Japan\\
$^{6}$Astronomical Institute, Tohoku University, Sendai, Miyagi 980-8578, Japan\\
$^{7}$Jet Propulsion Laboratory, California Institute of Technology, Pasadena, CA 91109, USA\\
\\
{\rm \copyright 2015. All rights reserved.}
}
\begin{document}

\date{Draft version \today}

\maketitle

\label{firstpage}

\begin{abstract}
The formation of supermassive stars (SMSs) is a potential pathway to seed 
supermassive black holes in the early universe. 
A critical issue for forming SMSs is stellar UV feedback, which
may limit the stellar mass growth via accretion. 
In this paper we study the evolution of an accreting SMS and its
UV emissivity under conditions of realistic variable accretion from 
a self-gravitating circumstellar disc.
First we conduct a 2D hydrodynamical simulation to follow the 
long-term protostellar accretion until the stellar mass exceeds 
$10^4~\msun$. 
The disc fragments due to gravitational instability, 
creating a number of small clumps that rapidly migrate inward 
to fall onto the star. 
The resulting accretion history is thus highly time-dependent:
short episodic accretion bursts are followed by longer, relative quiescent phases. 
We show that the circumstellar disc for the so-called direct collapse model
is more unstable and 
generates greater variability over shorter timescales than
normal Pop III cases.
We conduct a post-process stellar evolution calculation
using the obtained accretion history.
Our results show that, regardless of the strong variability
of the accretion rates, the stellar radius monotonically 
increases with almost constant effective temperature at 
$T_{\rm eff} \simeq 5000$~K as the stellar mass increases. 
The resulting UV feedback is too weak to hinder mass
accretion due to the low flux of stellar UV photons, thus
verifying our implicit assumption of no stellar feedback
during the hydrodynamic simulations.
The insensitivity of stellar evolution to variable accretion
is attributed to the fact that typical timescales 
of variability, $\lesssim 10^3$ years, are
too short to affect the stellar structure.
We argue that this evolution will continue until the SMS
eventually collapses to produce a massive black hole
by the general relativistic instability
after the stellar mass reaches $\gtrsim 10^5~\msun$.
\end{abstract}


\section{Introduction}
\label{sec:Introduction}

Supermassive black holes (SMBHs) with $\gtrsim 10^9~\msun$ 
already exist as early as at $z \gtrsim 6$ 
\citep[e.g.,][]{Fan2003aa,Fan2006aa,Willot2010aa,Mortlock:2011aa,Wu:2015aa}.
Formation of such SMBHs is a subject of intense study,
since little is known about the processes leading to the accumulation of 
such a tremendous amount 
of mass within a billion years.


A possible scenario for the early formation of SMBHs is the 
so-called direct collapse model: A massive seed BH with 
$\sim10^5~\msun$ increases its mass through accretion and mergers
to become a SMBH \citep[e.g.,][]{Bromm:2003uq}.
The massive seed BH can be formed by the collapse of a 
supermassive star (SMS) with $\sim10^5~\msun$.
The SMS is supposed to form under special conditions 
which allow very rapid protostellar accretion 
at an average rate of $\mdot \gtrsim 0.1~\msunyr$. 
The proposed possible conditions include, e.g.,  
gas clouds irradiated by strong photodissociating background 
\citep[e.g.,][]{Sugimura2014aa,Agarwal2015aa,Latif2015aa},
and dense shocked gas created through the formation of
protogalaxies \citep[e.g.,][]{IO12,Fernandez14,Inayoshi:2015aa}.


A potential obstacle for the formation of SMSs is
UV feedback from the protostar itself, which might halt stellar mass
growth via accretion \citep[e.g.,][]{MT08,HOYY11}. 
For normal Pop III cases with $\mdot \lesssim 0.04~\msunyr$, 
the protostar enters the so-called Kelvin-Helmholtz (KH)
contraction stage at some point. 
The stellar effective temperature rises as
the star contracts, so that UV feedback 
operates with a copious flux of UV photons. 
However, recent studies predict a
qualitatively different evolution at higher accretion rates
$\gtrsim 0.04~\msunyr$, i.e., for the direct collapse case
 \citep[e.g.,][]{Hosokawa:2013jk,Dominik13};
the protostar monotonically inflates with increasing
stellar mass and remains at a low effective
temperature $T_{\rm eff} \simeq 5000$~K
(named as the ``supergiant protostar'' stage).
At this low effective temperature, very few
ionizing photons are emitted by the star.


The above cited work assumed constant accretion
rates for simplicity. In reality, however, mass accretion 
occurring through self-gravitating circumstellar discs
should be highly time variable.
Such a disc often fragments to form clumps, 
some of which could be ejected, but most of which migrate
inward through the disc. Accretion rates are greatly enhanced 
for a short time when the clump falls onto the star.
The burst event is normally followed by a quiescent phase, 
where accretion almost ceases for a while. 
Numerical simulations predict that such episodic accretion
commonly appears both in present-day and primordial 
star formation \citep[e.g.,][]{VB06,Vorobyov2015aa,Machida2010aa,
Rowan12, Greif12, Vorobyov:2013lr,Hosokawa15}.  
Although still limited, recent studies have also
reported signatures of disc fragmentation 
for the direct collapse cases
\citep[e.g.,][]{Regan:2014rf,Becerra2015aa}. 


Variable accretion may allow the star to contract,
as the accretion rates can temporarily fall 
below $\simeq 0.04~\msunyr$, the critical value for 
bloating the star.
\citet[][hereafter SHYY15]{Sakurai:2015aa} have 
calculated the stellar evolution for various 
episodic accretion histories, assuming periodic high and low
states of accretion with a fixed mean value
at $0.1~\msunyr$.
They show that the star can contract if quiescent
phases with $\mdot \ll 0.04~\msunyr$ continue for 
$\Delta t_{\rm q} \gtrsim 10^3~{\rm yr}~[M_* / 500~\msun]^{1/2}$.
Once the star contracts, the stellar effective temperature 
and UV emissivity rapidly rise.
A longer quiescent phase results in a smaller
radius and thus a larger flux of UV photons, which potentially
allows an \hii region to expand around the star.


SHYY15 modeled accretion histories
with analytic functions to study how stellar evolution changes with 
different accretion histories in a controlled manner.
In this paper, as a next step, we investigate
stellar evolution with more realistic accretion
histories. 
We first conduct a 2D hydrodynamical simulation following
the protostellar accretion, using a central sink cell to 
monitor the accretion history onto the star.
The resulting accretion history is highly time-dependent
as the emerging circumstellar disc becomes
highly gravitationally unstable.
We then calculate stellar evolution using the obtained accretion history
as a post process and demonstrated that the star remains
in the bloated supergiant stage; the quiescent phases 
never last for $\Delta t_{\rm q} \gtrsim 10^3$ years
when $M_*\gtrsim 500~\msun$.


The rest of this paper is organized as follows. In \secref{sec:methods}, the 
methods of the hydrodynamical simulation and the stellar evolution calculation 
are briefly explained. In \secref{sec:results}, the results of the both calculations are presented.
In \secref{sec:conclusion and discussion}, we discuss the implications of
these results and summarize our conclusions.

\section{THE NUMERICAL APPROACH}
\label{sec:methods}
\subsection{2D hydrodynamical simulations}
\label{sec:sim methods}
Our model and method for numerically studying the gravitational collapse 
of primordial cores are presented in \citet{Vorobyov:2013lr}. 
Here, we briefly review the main concepts and appropriate
modifications for the formation of SMSs.
We follow the evolution of gravitationally unstable massive primordial
cores from the isolated pre-stellar stage into the star and
disc formation stages and terminate our simulations once about
50\% of the initial mass reservoir has been accreted onto the
star plus disc system. Once the disc is formed, it occupies
the innermost region of our numerical grid.
The dynamics of both the disc and envelope are followed
self-consistently on one global numerical grid, which ensures
correct mass infall rates onto the star plus disc system. This is an important
prerequisite for studying gravitational instability and fragmentation
in young circumstellar discs at all epochs 
\citep[e.g. ][]{Vorobyov2010aa,Machida2010aa,Vorobyov:2013lr}.


We introduce a sink cell at the inner boundary of our computational domain with a radius of 
$R_{\rm sc} = 110$~AU, and allow matter to freely flow into the sink cell. 
The radius of the sink cell is chosen to accommodate the maximum radius of
the growing star. 
In the early pre-stellar phase
of evolution, we monitor the mass accretion rate through the
sink cell and introduce a central point-mass object (representing
the forming star). 
In the subsequent evolution,
approximately 95\% of the accreted material is assumed to land directly onto
the star, whereas the rest is contained in the sink cell in order
to keep its density equal to the mean density of the gas in the
innermost 10--20 AU outside the sink cell. 


We solve the usual mass and momentum transport equations
written in the thin-disc approximation (see \secref{sec:sim results} for justification),
using a method of finite-differences with a time-explicit, operator-split procedure similar to that described
by \citet[][]{Stone1992aa} for their ZEUS-2D code. Advection is
performed using the third-order piecewise parabolic scheme \citep{Colella1984aa}. 
The gravitational acceleration includes contributions from
the central point-mass star (once formed), from material in
the sink cell ($r < R_{\rm sc}$), and from the self-gravity of the
circumstellar disc and envelope.

\begin{figure}
\centering
 \includegraphics[width=9cm]{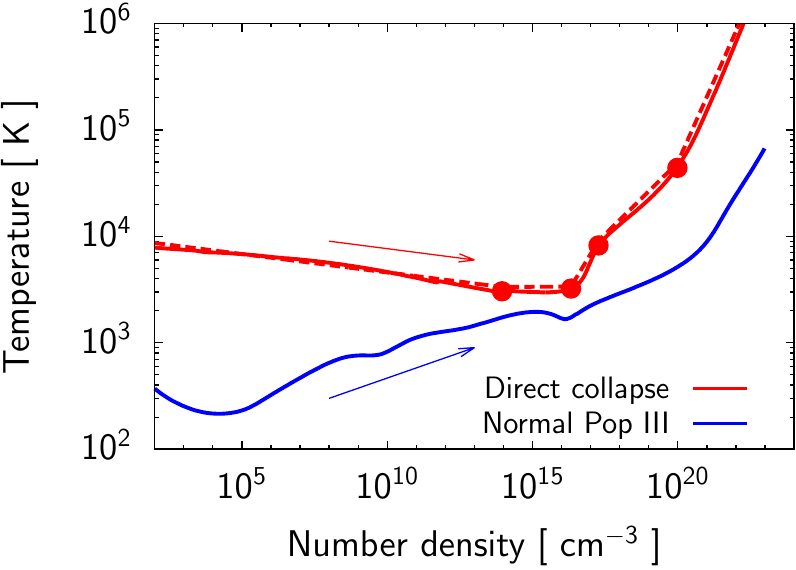}
 \caption{The temperature evolution of collapsing primordial gas as a function of the number density of hydrogen.
 The red line depicts the evolution of gas irradiated by a strong UV background which corresponds to the
 direct collapse case \citep[Figure 5a of][ $\lbrack{\rm M/H}\rbrack=-6$ ]{Omukai2008aa} considered here.
 The red dashed line portrays the piecewise polytropic fit used in our simulations (see text).
 The blue line displays the evolution of metal-free gas in the absence of UV background \citep[][]{Omukai2005aa}.
 }
 \label{fig:eos}
\end{figure}

The equations of mass and momentum transport are closed with a barotropic equation of state 
for the gas pressure $P$ of the following form
\begin{equation}
P_k = {\cal K} \rho^{\gamma_k} \prod_{i=1}^{k-1}\rho_{\rm{c},i}^{\gamma_i-\gamma_{i+1}},
\,\,\, \mathrm{for} \,\,\, \rho_{{\rm c},k-1} \le \rho < \rho_{{\rm c}, k} \, ,
\label{EoS}
\end{equation}
where ${\cal K}={\cal R} T/(\mu \rho^{\gamma_1-1})$,  $T=8000$~K is the initial gas temperature, 
$\cal R$ the universal gas constant, and $\mu=2.27$ the mean molecular weight of the primordial gas. 
Equation~(\ref{EoS}) is a piecewise fit to the
detailed thermal and chemical evolution during direct collapse calculated
by \citet{Omukai2008aa} using a one-zone model. 
Whereas the solid red line in Figure~\ref{fig:eos} portrays their 
exact solution, the red dashed line represents the
piecewise approximation used here. The index $k$ used in Table~\ref{table1} distinguishes between
the five individual components of the approximation. 
Also given in Table~\ref{table1} for each component $k$ are
the values of the various associated polytrope indices $\gamma_k$ and
the associated mass and number volume densities, 
$\rho_{\rm{c},k}$ and $n_{{\rm c},k}$, 
at which transitions between $k$ and $k+1$ occur (depicted by red dots in Figure~\ref{fig:eos}).
We note that 
when $k=1$ the product term is unity, and the pressure reduces to 
$P_1={\cal K} \rho^{\gamma_1}$. Moreover, ${\cal K}$ is approximately equal to the square 
of the initial sound speed $c_{\rm s}^2={\cal R} T/\mu$, because $\gamma_1=0.965 \approx 1.0$. 


In our simulations, the corresponding form of the barotropic relation used in the code is
\begin{equation}
{\cal P}_k = {\cal K} \Sigma^{\gamma_k} \prod_{i=1}^{k-1}\Sigma_{\rm{c},i}^{\gamma_i-\gamma_{i+1}},
\,\,\, \mathrm{for} \,\,\,  \Sigma_{{\rm c},k-1} \le \Sigma < \Sigma_{{\rm c}, k} \, ,
\label{eos}
\end{equation}
where $\cal P$ is the vertically integrated gas pressure; the transition surface and volume mass density are related to one another through the instantaneous local scale height $Z$ at each point in the disc via $\Sigma_{{\rm c},i}=2Z \rho_{{\rm c},i}$. The scale height $Z$ is calculated assuming a
local hydrostatic balance in the gravitational field of both the star and the disc \citep{Vorobyov2009}.


\begin{table}
\begin{center}
\caption{Parameters of the Barotropic Relation}
\label{table1}
\renewcommand{\arraystretch}{1.5}
\begin{tabular*}{\columnwidth}{ @{\extracolsep{\fill}} c c c c }
\hline \hline
$k$ & $\gamma_i$ & $\rho_{\rm{c},i}$ & $n_{\rm{c},i}$ \\
\hspace{1cm} & & (g~cm$^{-3}$)   & (cm$^{-3}$) \\ [0.5ex]
\hline \\ [-2.0ex]
1 & 0.965 & $3.38{\times}10^{-10}$ & $8.92{\times}10^{13}$ \\
2 & 1.002 & $8.037{\times}10^{-8}$ & $2.12{\times}10^{16}$ \\
3 & 1.456 & $7.089{\times}10^{-7}$ & $1.87{\times}10^{17}$ \\
4 & 1.269 & $3.673{\times}10^{-4}$ & $9.69{\times}10^{19}$ \\
5 & 1.614 & --- & --- \\ [1.0ex]
\hline
\end{tabular*}
\end{center}
\end{table}


The initial gas surface density $\Sigma$ and angular velocity $\Omega$ profiles 
are similar to those that were considered in the context of 
normal population III star formation \citep{Vorobyov:2013lr}
\begin{equation}
\label{ic1}
\Sigma = {r_0 \Sigma_0 \over \sqrt{r^2+r_0^2}}\:,
\end{equation}
\begin{equation}
\label{ic2}
\Omega = 2 \Omega_0 \left( {r_0\over r}\right)^2 \left[\sqrt{1+\left({r\over r_0}\right)^2} -1\right].
\end{equation}
The radial profile of $\Sigma$ is an integrated form of a Bonnor-Ebert sphere, while that of 
$\Omega$ is the expected differential rotation profile to accompany Equation \eqref{ic1} for a core contracting from near-uniform initial conditions \citep{Basu1997aa}. 
The parameters $\Omega_0=7.22$~km~s$^{-1}$~pc$^{-1}$, $\Sigma_0=7.63$~g~cm$^{-2}$, and $r_0=0.154$~pc, are the central angular velocity, central gas surface density, and the radius of a central near-constant-density plateau, respectively. 
They are chosen to create a gravitationally unstable core with initial 
mass $M_{\rm c}=26240~M_\odot$ and ratio of rotational to gravitational energy 
$\beta=1.96\times 10^{-2}$. 
Although the initial cloud mass is somewhat lower than that assumed 
for the direct-collapse model, where SMSs exceeding $10^5~\msun$ 
may finally form, this is sufficient to allow us to follow the evolution for
the first $\sim 10^5$ years of protostellar accretion.
Below we compare the resulting evolution with that discussed in 
\citet{Vorobyov:2013lr}, who have followed the 
evolution for a similar duration but
for the normal Pop III case. 


Numerical simulations are run on a polar coordinate ($r,\phi$) grid with 
$512 \times 512$ spatial zones. The radial points are logarithmically spaced, allowing improved
numerical resolution of the inner grid, where the disc forms and evolves. The innermost cell 
outside the central sink has a radius $R_{\rm sc} + 1.6$~AU
and the radial and azimuthal resolution are about 
14~AU at a radius of 1000 AU and 70~AU at a distance of 5000~AU. 
This resolution is sufficient to fulfill the Truelove criterion, 
which states that the local Jeans length must be resolved by at least four numerical cells 
\citep{Truelove1997aa}. Indeed, the Jeans length of a thin
self-gravitating disc can be written as \citep{Vorobyov2013bb}
\begin{equation}
\label{RJeans}
 R_{\rm J} = { \overline{c_{\rm s}}^2 \over G \overline{\Sigma} }  \;.
\end{equation}
For the mean surface density of $\overline{\Sigma}\,{\approx}\,500~\mbox{g~cm}^{-2}$ 
and mean temperature $\overline{T}\,{\approx}\,7500~\mbox{K}$, typical for our disc at 
$r=1000-5000~\mbox{AU}$ (see Fig.~\ref{fig:radial}), the corresponding Jeans length is 
$R_{\rm J}{\approx}550~\mbox{AU}$ and is resolved by roughly 40 grid zones at 1000~AU and 8 grid zones
at 5000~AU in each direction $(r,\phi)$.


\subsection{Stellar evolution calculations}
\label{sec:se methods}
We use the stellar evolution code STELLAR originally
developed by \citet[][]{Yorke:2008rz}, which has been 
also used in SHYY15.
Since the detalied description of the code is in
SHYY15, we here briefly explain the main 
features of the code.


The code solves the basic equations of stellar evolution,
including effects of mass accretion under the assumption of spherical symmetry.
Nuclear reactions are considered up to helium burning (3$\alpha$ and \{CNO\} + He).
Energy transport by convection is modeled by mixing
length theory. 


We use a grey atmosphere boundary condition 
for the stellar surface layer, where the accreted gas accumulates.
The gas mass $\mdot\Delta t$ is added to 
the outermost grid cell each time step, 
where $\mdot$ is the accretion rate and $\Delta t$ is the stellar evolution time step. 
The physical quantities of the accreted gas are 
assumed to be the same as those in the outermost grid point. 
This approximates an extreme case where the accreting gas slowly 
approaches the star and has time to adjust thermally to the stellar
surface, but is not always the case because the gas 
can accrete with more thermal energy.
We take this into account by parametrizing the fraction of 
accretion luminosity deposited onto the stellar surface,
\begin{equation}
\eta \equiv \frac{L_{*,\mathrm{acc}}}{L_{\rm acc}}
= L_{*,\mathrm{acc}}  
  \left( \frac{G M_* \dot{M}_*}{R_*} \right)^{-1}
~,
\end{equation}
where $L_{*,\mathrm{acc}}$ is the portion of accretion luminosity which contributes 
directly to the stellar luminosity and thus affects the stellar structure.
As described in \citet[][]{Hosokawa:2013jk}, the parameter $\eta$ has a little 
effect on stellar evolution for high accretion rates $\gtrsim0.1~\msunyr$ aside 
from the earliest accretion phases.
We assume $\eta=0.1$ in our calculations.


The initial condition is a polytrope star of $2~\msun$ with polytropic index 
$n=1.5$, which well approximates a fully convective star. Before commencing
with mass growth, the polytrope is allowed to first relax to a fully converged stellar
model using STELLAR.
The composition of the gas is assumed to be pristine with $X=0.72$
and $Y=0.28$. The accretion history is taken from the 2D hydrodynamical 
simulation (see \secref{sec:sim results}).

\section{Results}
\label{sec:results}
\subsection{Episodic mass accretion with self-gravitating discs}
\label{sec:sim results}

\begin{figure}
\centering
 \includegraphics[width=9cm]{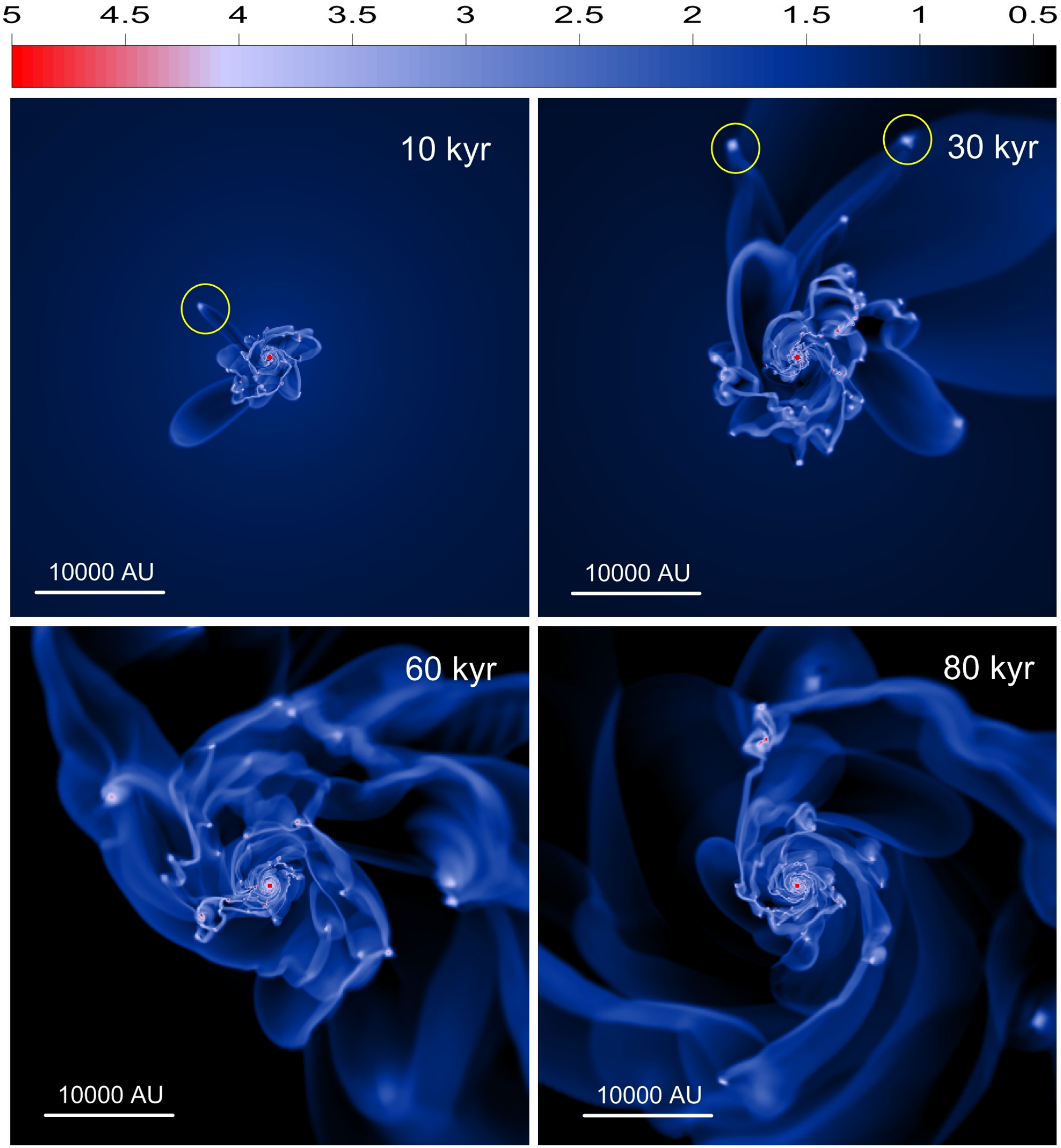}
 \caption{Images of the gas surface density in the inner $20000 \times 20000$ AU$^2$ box 
 showing the evolution of the disc around the SMS star 
 (represented schematically as a red circle in the coordinate center). The elapsed time since
 the formation of the star is indicated in each panel. {\bf\rm Yellow circles indicate the fragments 
 that are ejected from the disc.}
 The scale bar is in g~cm$^{-2}$.
 }
 \label{fig:surfdens}
\end{figure}


We start by describing the evolution of a circumstellar disc formed 
as a result of the gravitational collapse of our massive primordial core.
Figure~\ref{fig:surfdens} presents the time evolution of the gas surface density in the 
inner $20000\times20000$~AU$^2$ box. The total computational region is 
about $10^2$ times 
larger in area
than shown in Figure~\ref{fig:surfdens}.
The elapsed time since the formation of the star (schematically shown by red circles in the coordinate
center) is shown in each panel. Evidently, the disc is strongly gravitationally unstable and 
quickly breaks into giant spiral arcs. Gravitationally bound and pressure supported clumps form within these arcs via gravitational fragmentation.


In the absence of sink particles, tracking fragments during numerical simulations (on the fly)
in grid-based codes becomes a challenging task.
Therefore, we analyze the properties of the fragments in a post-process mode,
using the method described in detail 
in \citet{Vorobyov:2013lr} in the context of discs around Population III stars. 
The algorithm is based on two conditions: The first one dictates 
that the fragment must be pressure supported, with a negative
pressure gradient with respect to the center of the fragment. The
second condition requires that the fragment be held together by
gravity, with a positive gravitational potential gradient from the 
deepest potential well located at the center of the fragment.

\begin{figure}
\centering
 \includegraphics[width=9cm]{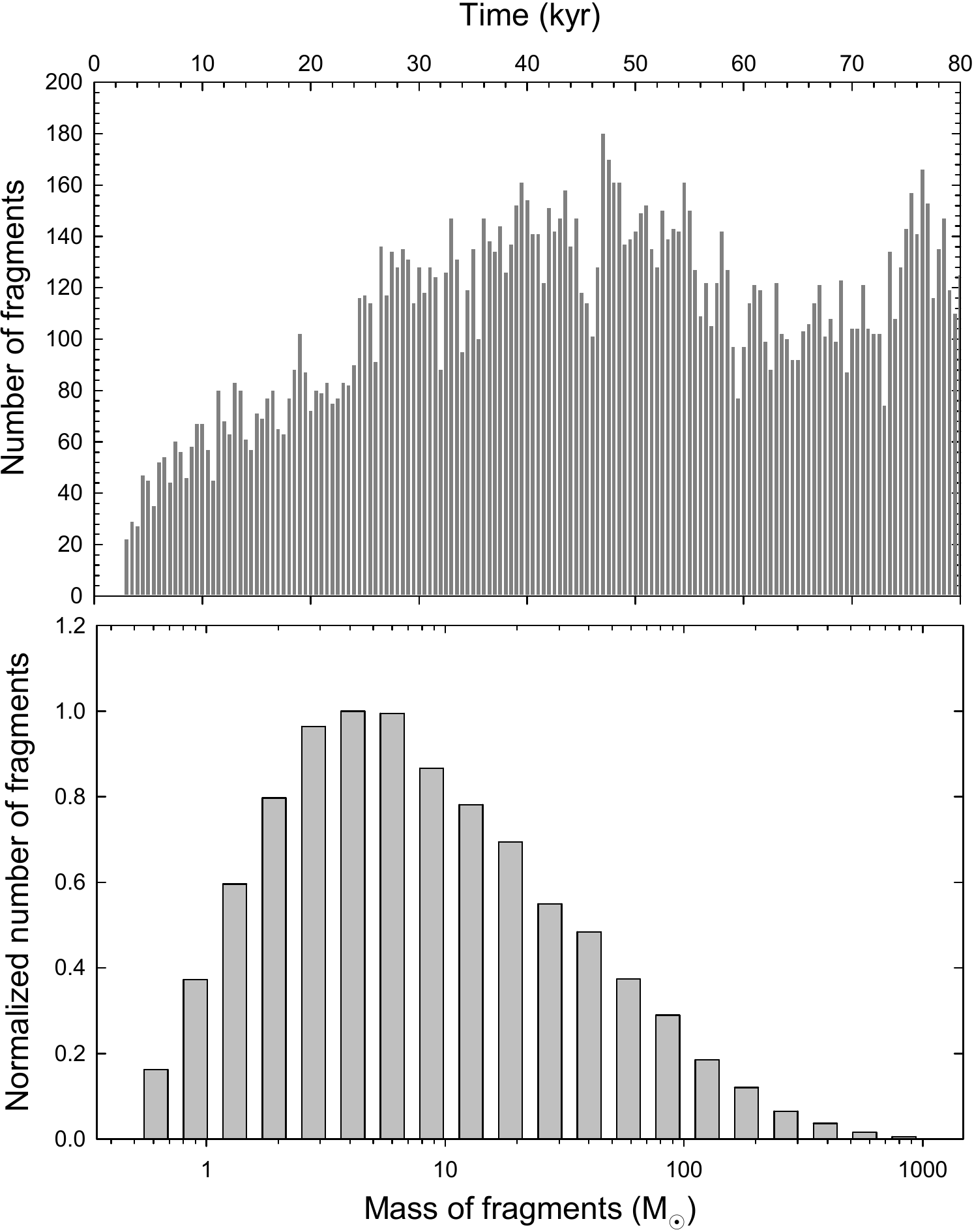}
 \caption{{\bf Top}. Normalized number of fragments in the disc as a function of the elapsed time since
 the formation of the SMS star. {\bf Bottom.} The distribution function of masses of the fragments formed
 in the disc at all times.}
 \label{fig:fragments}
\end{figure}

The top panel in Figure~\ref{fig:fragments} presents the number 
of fragments in the disc as a function of time. 
The number of fragments grows with time from a few tens immediately after the formation of the central star to more 
than a hundred by the end of our simulations. We note that the growth is not steady but 
is characterized by both increases and decreases in the number of fragments, implying 
that fragments can be both created and destroyed
or otherwise lost by the disc. One such loss channel is accretion of the fragments onto the star 
caused by the loss of angular momentum due to gravitational interaction with other fragments 
and spiral arcs. This phenomenon is described in detail in \citet{VB06,Vorobyov2010aa} in the context
of present-day star formation and in \citet{Vorobyov:2013lr} in application to the formation of Pop III
stars. 


\begin{figure}
\centering
\includegraphics[width=8cm]{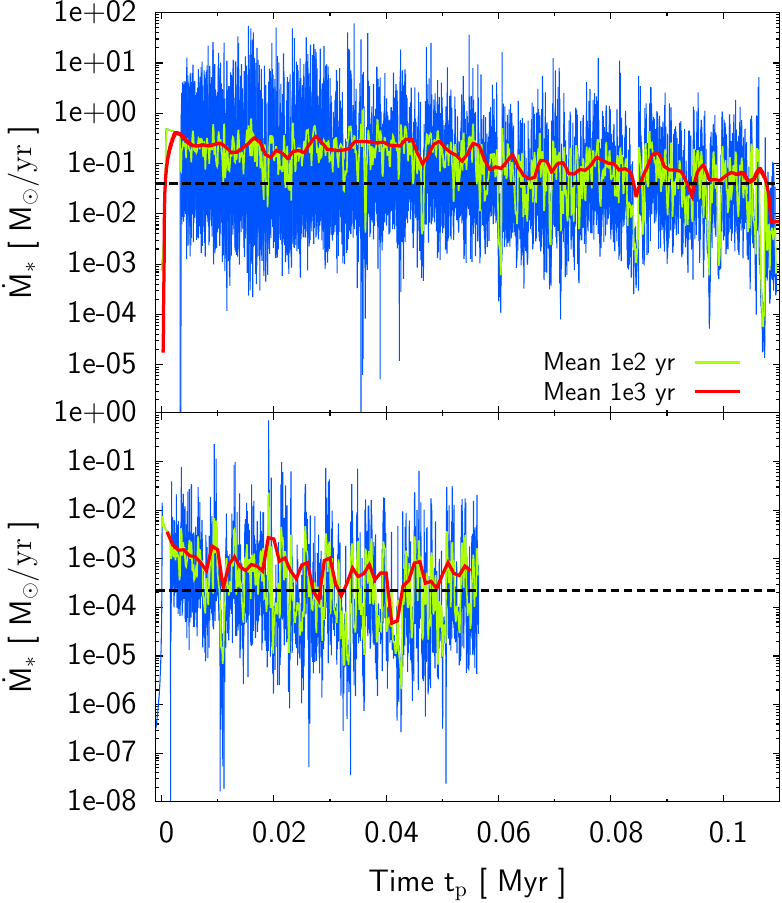}
 \caption{ Comparisons of the accretion histories between 
the direct collapse and the normal Pop III cases.
The origin of the time axis is the epoch of
formation of the protostar.
{\sl Top panel:} the accretion history for the direct collapse case
obtained by the 2D simulation.
The blue line depicts the original accretion history without time averaging, 
and the red and green lines denote the time-averaged
histories with bins of $10^3$ and $100$ years. 
{\sl Bottom panel:} the accretion history for the normal Pop III 
case taken from \citet[][]{Vorobyov:2013lr}. 
The different colors have the same meanings as in 
the top panel. 
In each panel, the black dashed line indicates the threshold 
accretion rate below which the accretion is in the quiescent phase (see text).}
 \label{fig:acc_average}
\end{figure}

The number of fragments and the rate of their formation in our model disc is 
roughly a factor of 10 greater than was found in discs around present-day stars 
or normal Pop III stars \citep{Vorobyov:2013aa,Vorobyov2015aa}.
This increase in the number of fragments is a consequence of the specific temperature
vs. density relation typical for the direct collapse case. As the red
line in Figure~\ref{fig:eos} demonstrates, 
there is a wide range of gas densities in the disc, $\le 10^{16}$~cm$^{-3}$, for which the gas 
temperature decreases as the density increases.    Any compression therefore creates 
a positive feedback for gravitational instability -- an increase in density leads 
to a decrease in temperature, which in 
turn promotes further compression and, ultimately, fragmentation. 
An increased number of fragments leads to much higher accretion 
burst activity in SMS stars than for present-day or Pop III stars  
(see \figref{fig:acc_average} and Section 3.2 below).


The bottom panel in Figure~\ref{fig:fragments} shows the normalized
distribution function of masses of the fragments. We calculate 
the distribution function using all fragments identified in the top panel. 
Since the time sampling in the top panel is 500~yr, 
some long-lived fragments may be duplicated so that the calculated distribution function 
will be skewed towards long-lived fragments.
The fragments vary 
in mass from a fraction of a solar mass to several hundred solar masses. 
Our simulations indicate that some of the fragments may be ejected out of
the disc via multi-body interactions.
Several such candidates are highlighted in Figure~\ref{fig:surfdens}
by the yellow circles.
If they can acquire sufficient velocities
to escape the gravitational pull of the disc and the star, this may represent an
interesting gateway for the formation of freely-floating primordial stars. 


It is interesting to compare the variable accretion histories
among the direct collapse case and the normal Pop III
case studied by \citet{Vorobyov:2013lr}, focusing on the duration of quiescent phases of the accretion $\Delta t_{\rm q}$.
In order to quantify the duration of quiescent phases,
we henceforth define the quiescent phases for the direct collapse case as 
phases when accretion rates are below $0.04~\msunyr$,
the critical rate above which the star enters the "supergiant" phase (see Section 1).


For the normal Pop III case, 
we define the quiescent phases as phases when the accretion rate falls below one tenth of the 
mean value. The definition is an analogy to that for the direct collapse case: 
the critical accretion rate below which accretion rates become quiescent is 
$\sim0.1\times$ the mean accretion rate for the direct collapse case.
In the following, we estimate $\Delta t_{\rm q}$ from the accretion histories for each case.
In \figref{fig:acc_average}, the evolution of the accretion histories is shown 
for the direct collapse case (top panel) and the PopIII case (bottom panel).  
The accretion history is shown as a function of the
elapsed time since the formation of a protostar $t_{\rm p}$.
The blue line shows the original accretion history, while the red and green lines indicate 
the accretion histories averaged over $\Delta t_{\rm bin}=10^3$ and $100$ years, respectively. 
In each panel, the black dashed line is the threshold accretion rate
$\dot{M}_{\rm th}$ below which the accretion is deemed quiescent. 
When averaging over $\Delta t_{\rm bin}$, all
variations shorter than $\Delta t_{\rm bin}$ are smoothed. 
Thus, the duration of quiescent phases
$\Delta t_{\rm q}$ will be less than $\Delta t_{\rm bin}$,
if the accretion rate averaged over $\Delta t_{\rm bin}$  exceeds $\dot{M}_{\rm th}$.
According to the above consideration, the typical duration of quiescent 
phases in the direct collapse case is much smaller than $10^3$~yr 
for $t_{\rm p} \lesssim 0.06~\Myr$, because the accretion rate
averaged over $10^3$~yr never falls below the 
threshold value and the accretion rate
averaged over $10^2$~yr seldom (and only for short periods of time $< 10^3$~yr)
drops below the threshold value.
The relatively long quiescent phases with $\Delta t_{\rm q} \sim 10^3$
years only appear in the late accretion stages for $t_{\rm p} \gtrsim 0.06~\Myr$,
because of the gradual depletion of the envelope mass with time and 
associated weakening of gravitational fragmentation in the disc.
By contrast, the quiescent phases for the normal Pop III case
are much longer, $\Delta t_{\rm q} \gtrsim 10^3$ years 
up to the end of the calculation, because the accretion rate
averaged over both $10^2$~yr and $10^3$~yr falls below the threshold value
for significant periods of time.

\begin{figure}
\centering
 \includegraphics[width=9cm]{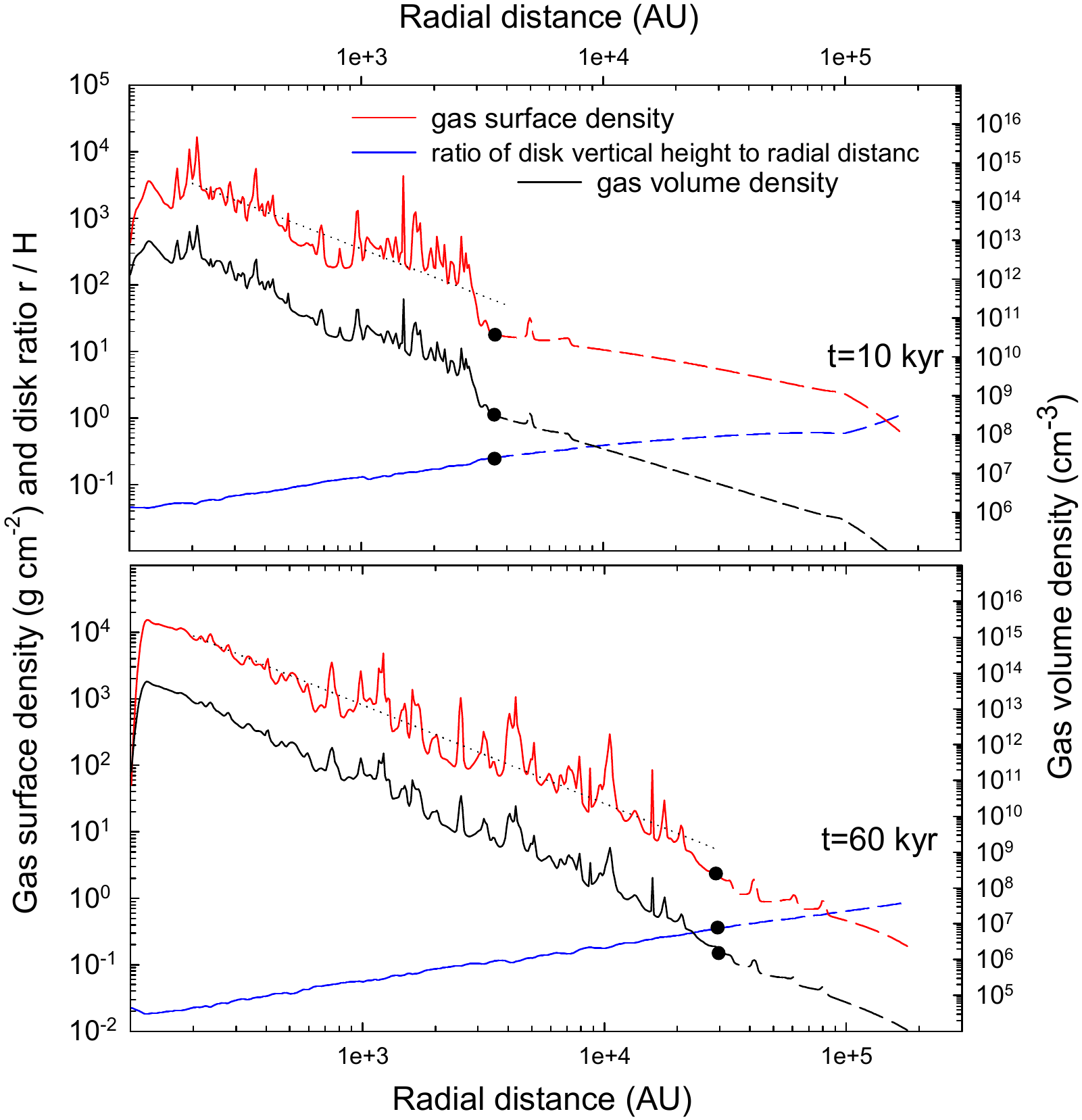}
 \caption{
 Azimuthally averaged gas surface 
 and volume density profiles 
 (red and black lines respectively) at two different evolutionary 
 times after the formation of the star. The black circles mark the position of the disc outer edge. 
 The black dotted lines provide the least-squares fits to 
 the gas surface density profiles, which follow approximately an $r^{-1.5}$ law. 
 The blue line is the ratio of the disc vertical scale height
 to the radial distance from the star. 
 The outside regions of the disc are represented by the dashed lines.}
 \label{fig:radial}
\end{figure}

We attribute the above difference of $\Delta t_{\rm q}$ to the 
different intervals of disc fragmentation. As previously described,
the disc in the direct collapse case is more unstable than in
the normal Pop III case. This means that disc fragmentation is
also more frequent. 
A greater rate of fragments infalling onto the star
results in shorter durations of quiescent phases. 
The key quantity for the gravitational stability of the disc 
is the so-called Toomre $Q$ parameter 
\citep{Toomre1964}. 
It is known that, for the accretion stage of star formation, the
$Q$ parameter is approximately given by 
$Q\sim \mathcal{O}(0.1-1)\times(T_{\rm disc}/T_{\rm env})^{3/2}$,
where $T_{\rm disc}$ and $T_{\rm env}$ are the temperatures
of the disc and surrounding envelope 
\citep[e.g.,][]{Kratter2010aa,Tanaka2014aa}. 
For $T_{\rm disc} < T_{\rm env}$, the disc 
becomes more gravitationally unstable at a smaller $Q$ value. 
As can be surmised from \figref{fig:eos}, such a temperature imbalance can occur
for the direct collapse case. 
In our 2D simulation, the number density at the boundary
between the disc and envelope is $\sim 10^{6-9}~{\rm cm}^{-3}$
(see \figref{fig:radial}).
Since the temperature is a decreasing function of the density
for $n_{\rm H} \lesssim 10^{16}~{\rm cm}^{-3}$, the disc 
temperature is slightly lower than the envelope temperature. 
By contrast, the temperature imbalance between disc and envelope is opposite for the
normal Pop III case, for which the temperature increases
with increasing the density. 
This results in a less unstable disc, which explains the
longer quiescent duration $\Delta t_{\rm q}$
seen in \figref{fig:acc_average}.


Finally, in Figure~\ref{fig:radial} we present the azimuthally averaged 
gas surface and volume density as a function of distance from the star for
two evolutionary times: 
one corresponding to the early evolution (10~kyr) and the other to the late evolution
(60~kyr). Whereas low-mass fragments and spiral arcs are washed out by azimuthal 
averaging, the existence of massive fragments in the disc becomes apparent
through the multiple short-wavelength peaks in the surface density distribution.
The black circles mark the position of the disc outer edge.
The positions of the black circles is calculated visually taking into account that 
fragments (manifested by sharp local peaks in the density profiles) 
form within the disc, and not within the infalling parental core. 
The black circles therefore separate the regions with varying (disc) and smooth (core) 
density profiles. We note that applying more sophisticated disc tracking mechanisms 
\citep[e.g.,][]{Dunham2014} has proven difficult for our highly unstable and fragmenting discs. Moreover,
some fragments may be later scattered outside the disc owing to multi-body
interactions, which would artificially increase our disc size estimates if we use the 
latter disc identification mechanism.


The black dotted lines are the least square fits to the surface density distribution 
in the disc. The corresponding relations at $t=10$~kyr and $t=60$~kyr are:
\begin{eqnarray}
\Sigma\left[ \rm{g} \over \rm{cm}^{2} \right] &=& 10^{6.7\pm0.2} \left( {r \over \rm{AU}} \right)^{-1.4\pm0.06}, \\
\Sigma\left[ \rm{g} \over \rm{cm}^{2} \right]  &=& 10^{7.4\pm0.1} \left( {r \over \rm{AU}} \right)^{-1.5\pm0.02}.
\end{eqnarray}
The disc surface density follows approximately an $r^{-1.5}$ law, also 
typical for self-gravitating discs  around Pop III stars and in the local Universe 
(Vorobyov et al. 2013). The blue lines depict
the ratio of disc vertical scale height to distance from the star. 
Evidently, this quantity stays well below unity everywhere in the disc,
justifying the use of the thin-disc approximation. 


\subsection{Stellar evolution under episodic accretion}
\label{sec:se results}

\begin{figure}
\centering
 \includegraphics[width=7.9cm]{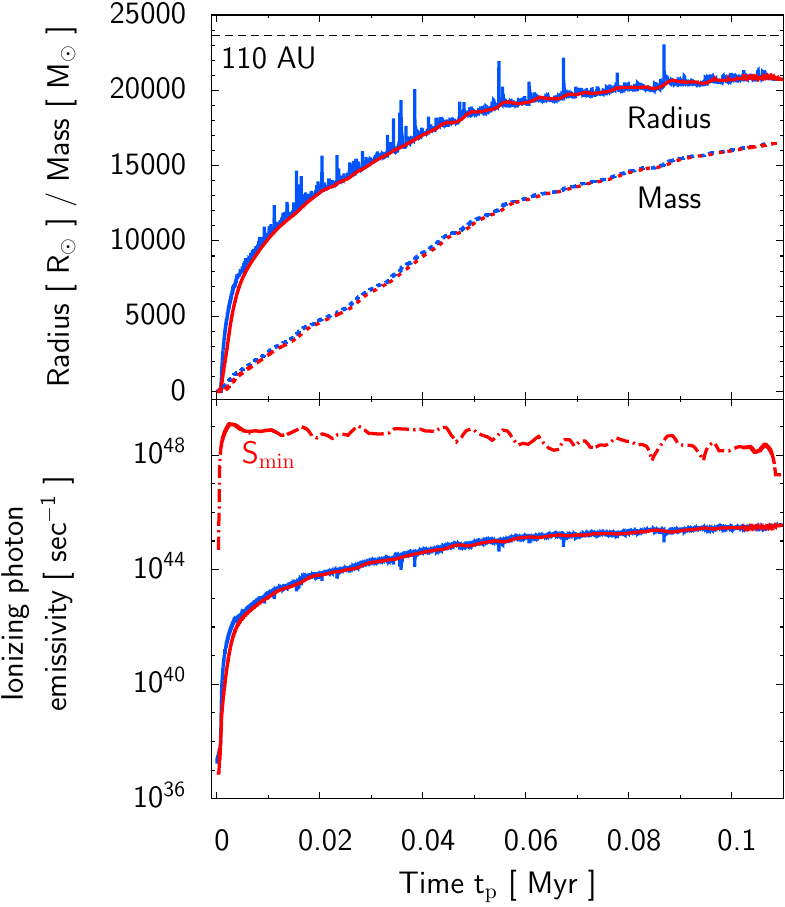}
 \caption{
 The time evolution of stellar mass and 
radius (top panel) and ionizing photon emissivity (bottom panel). 
The origin of the time axis is the epoch of
formation of the protostar.
The blue lines show the evolution without time averaging 
of the accretion rate (see \figref{fig:acc_average}). 
The red lines show the evolution with the
accretion rates averaged with time bins $10^3~\yr$.
The black dashed line in the top panel indicates 
the sink radius of 110~AU.
In the bottom panel, the dot-dashed line represents the critical
value above which an \hii region will appear 
(see \secref{sec:conclusion and discussion}).}
 \label{fig:MDOTMRx}
\end{figure}


We now consider the evolution of the central protostar under the variable
accretion history (blue curve in the top panel of \figref{fig:acc_average})
obtained from the hydrodynamical simulation 
described in \secref{sec:sim results}. 
\figref{fig:MDOTMRx} shows the evolution of the 
stellar mass and radius and the ionizing photon emissivity.
As seen in the top panel of \figref{fig:acc_average},  
the mean accretion rate approximately spans
$\simeq 0.1-0.3~\msunyr$, which is expected for the direct collapse model.
Despite the frequent drops of the accretion rate below 
$0.04~\msunyr$, which could in principle allow stellar contraction, 
the protostar's radius increases almost monotonically 
until the end of the calculation. 
The stellar radius reaches 100~AU at the end of the simulation, 
only slightly below the sink radius of 110~AU. 
This evolution is similar to that for constant accretion with rates of
$\gtrsim0.1~\msunyr$ (e.g., Figure 1 of SHYY15). 
Not shown in the figure is the fact that the effective temperature remains 
almost constant at $\simeq 5000$~K
due to the very sensitive temperature-dependence of H$^-$ opacity. 
The ionizing photon emissivity therefore remains insufficient 
for creating an \hii region. 


SHYY15 concluded that, in order for the star to contract and leave the supergiant
stage, the quiescent phase has to be longer than at least $10^3$ years 
for $M_*\gtrsim500~\msun$. 
This explains the absence of contraction during the quiescent 
phases of our current simulations.
SHYY15 derive the critical duration of $10^3$ years at $500~\msun$
by considering the typical KH timescale of supergiant protostars. 
As shown in previous studies \citep[e.g.,][]{Hosokawa:2013jk},
the mass distribution in the interior of the bloated protostar is 
highly inhomogeneous; only a surface layer with a 
very small fraction
of the total mass largely inflates to cover most of the stellar
radius. SHYY15 have thus used the KH timescale only for the bloated
surface layer $t_{\rm KH,surf}$ (see their equation 10), instead of 
the usual global definition $t_{\rm KH} \equiv G M_*^2/R_* L_*$. 
SHYY15 show that the surface KH timescale is approximately 
\begin{equation}
t_{\rm KH,surf} \simeq 10 t_{\rm KH} \simeq
10^3~{\rm yr}~\left(\frac{M_*}{500~\msun}\right)^{1/2} \;.
\end{equation}
Since the quiescent phases 
found in our simulation are all
shorter than this timescale, the inflated surface layer
does not have enough time to contract by radiating away thermal energy.
Even for the later stage $t_{\rm p} \gtrsim 0.06$~Myr 
when somewhat longer quiescent phases $\Delta t_{\rm q} \sim 10^3$ 
years appear, the star does not contract because $t_{\rm KH, surf}$ has
also increased to $\gtrsim 10^3$ years with increasing stellar mass.


In order to evaluate the strength of UV feedback, 
we use the critical value of ionizing photon emissivity 
$S_{\rm min} =\dot{M}/\mu m_{\rm H}$: the value to ionize all of the atoms 
infalling onto the star once \citep[see also SHYY15 and the discussion
of \hii region ``squelching'' by][]{Yorke1986}.
$S_{\rm min}$ is the lower limit to create an \hii region in spherical geometry,
because, in reality, additional UV photons are needed to ionize
recombined atoms.
Of course, the condition of ``squelching'' has to be fulfilled in all directions
in order to entirely prevent an \hii region from forming, and it is possible that 
a bipolar \hii region could expand into the directions, from which little accretion 
occurs, but the extremely low UV flux 
ensures that the
envelope can only be minimally affected by ionization. 
The bottom panel of \figref{fig:MDOTMRx} also shows
the evolution of $S_{\rm min}$, for which the accretion history
averaged over $10^3$ years is used.
Since the UV emissivity for the current case is always
much smaller than $S_{\rm min}$, no significant \hii egion will
appear to disturb the mass accretion.

\section{Conclusion and Discussion}
\label{sec:conclusion and discussion}
We have investigated the evolution of 
an accreting SMS and the resulting
stellar UV emissivity resulting from realistic variable mass accretion rates
generated by a self-gravitating circumstellar disc.
The numerical hydrodynamics simulation gives evidence of very dynamic features of the 
disc and protostellar
accretion; the disc readily fragments and the fragments then migrate
inward to fall onto the star. The resulting accretion history 
is highly time-dependent, characterized by a number of short episodic
accretion bursts followed by somewhat longer quiescent phases.
Despite the strong variability of the accretion rate, 
the resulting stellar evolution is quite similar to
that for constant accretion rates: namely, the stellar radius
increases monotonically with increasing stellar mass. 
The effective temperature is almost constant at $\simeq 5000$~K,
a temperature at which the star produces a negligible flux of UV photons. 
The absence of KH contraction during quiescent accretion phases is
due to their short duration, $\Delta t_{\rm q} \lesssim 10^3$
years. 
As shown by SHYY15, this is shorter than the local KH timescale
for the bloated stellar surface layer for $M_* \gtrsim 500~\msun$. 
Since the surface layer can only inefficiently radiate away thermal energy in such a short time, 
the star does not contract and leave the supergiant protostar stage.


In the current study, we have computed the evolution 
until the stellar mass reaches $\simeq 1.6 \times 10^4~\msun$
(Fig.~\ref{fig:MDOTMRx}).
As mentioned in Section~\ref{sec:sim methods}, this is mostly 
due to our adopted initial conditions, in particular, the limited
cloud mass of $\simeq 2.6 \times 10^4~\msun$. 
It would be possible to simulate a longer evolution up to higher stellar masses
assuming a higher initial cloud mass, but we focused on capturing the variability 
during the early stages of protostellar accretion, because
the surface KH timescale is shorter for 
lower stellar masses
$t_{\rm KH,surf} \sim 10^3~{\rm yr}~[M_*/10^3~\msun]^{1/2}$.


We note that with the default sink size $110$~AU,
it is not until the stellar mass reaches 
$\simeq 1000~\msun$ that the disc first appears.
\footnote{
The smaller the sink cell, the earlier the disc forms --
disc formation occurs when the infalling material 
first hits its centrifugal barrier. With our chosen initial conditions,
the lower angular momentum material, which
hits its centrifugal barrier closer to the central star, arrives
earlier.
}
A test case with a $70$~AU sink
shows that the disc and resulting accretion variability 
appear earlier for $M_* \gtrsim 700~\msun$, 
for which $\Delta t_{\rm q}$ is still lower than 100 years
as in the default case. 
Thus, we do not currently expect stellar contraction to occur in 
such early stages, a fact to be checked by future simulations.
Note that we have not adopted the smaller sink,
because the radius of the inflating SMS would soon exceed the sink size. 


Although the top panel of Figure~\ref{fig:acc_average} shows that the duration of 
some quiescent phases becomes longer for $t_{\rm p} \gtrsim 0.06~\Myr$, 
we expect that this is mostly due to  
the gradual depletion of the accretion envelope 
(Section~\ref{sec:se results}).
For more realistic cases of direct collapse, whereby significantly more massive clouds form 
in so-called atomic-cooling haloes, this mass depletion would be postponed to even later times.
However, the duration of quiescent phases may increase for other
reasons than envelope depletion.
Following the discussion in SHYY15, 
the typical time lag between accretion bursts 
is proportional to the fragmentation timescale 
$t_{\rm frag} = M_{\rm d}/\dot{M}_{\rm d}$, where $M_{\rm d}$
is the disc mass and $\dot{M}_{\rm d}$ is the gas infall rate onto
the disc \citep[also see][]{Vorobyov:2013aa}.
From the fact that the disc mass increases in proportion to
the stellar mass $M_{\rm d} \simeq 0.5~M_*$ in our simulation, one may surmise that
the fragmentation timescale also increases for $M_* \gtrsim 10^4~\msun$.
Since the typical duration of the quiescent phase is 
$\Delta t_{\rm q} \simeq 10^2$ years for $M_* \sim 10^4~\msun$
(see top panel of \figref{fig:acc_average}), $\Delta t_{\rm q}$ should reach
$\sim 10^3$ years for $M_* \sim 10^5~\msun$.
Nevertheless, this is still ten times shorter than the surface KH
timescale $t_{\rm KH,surf}$, which also increases with
stellar mass. 
Therefore, we do not expect significant stellar contraction for higher SMS masses,
until the stellar mass exceeds $\sim 10^5~\msun$. At this point, the SMS
is expected to collapse and produce a massive BH via the general relativistic
instability.

\section{Acknowledgements}
YS thanks the support from Advanced Leading Graduate Course for Photon Science.
EIV acknowledges the support from the Russian Ministry
of Education and Science Grant 3.961.2014/K.
This work was financially supported 
by Grant-in-Aid for JSPS Fellows
(15H00776: YS), by JSPS-OEAD short-term
research grant \#RC (21402003: EIV), and by the Grants-in-Aid for Basic Research
by the Ministry of Education, Science and Culture of Japan 
(25800102, 15H00776: TH, 25287040: KO, 25287050: NY).
Portions of this work were conducted at the Jet Propulsion Laboratory,
California Institute of Technology, operating under a contract with 
the National Aeronautics and Space Administration (NASA).
Numerical simulations were done 
on the Vienna Scientific Cluster (VSC-2), Atlantic Computational  Excellence Network (ACEnet), 
and Shared Hierarchical Academic Research Computing Network (SHARCNET).
The stellar evolution calculations were partly conducted on a PC cluster 
at the Center for Computational Astrophysics, National Astronomical Observatory of Japan.

\bibliography{mn-jour,ms}

\end{document}